%% file: main.tex
\documentclass[runningheads,a4paper]{llncs}

\makeatletter
\newcommand{\@chapapp}{\relax}%
\makeatother

\usepackage{etoolbox}

\usepackage[utf8]{inputenc}

\usepackage{algpseudocode}
\usepackage{algorithm}

\input{styles.tex}			

\input{macro.tex}			

\newcommand{\mytitle}{An analysis of Bitcoin \opreturn metadata}

\begin{document}

\mainmatter

\titlerunning{\mytitle}
\title{\mytitle}

\author{Massimo Bartoletti \and Livio Pompianu}
\authorrunning{Bartoletti M., Pompianu, L.}
\tocauthor{Massimo Bartoletti and Livio Pompianu}
\institute{Universit\`a degli Studi di Cagliari, Cagliari, Italy \\ \email{\{bart,livio.pompianu\}@unica.it}}

\maketitle

\input{abstract.tex}
\input{intro.tex}
\input{background.tex}
\input{methodology.tex}
\input{qualitative.tex}
\input{analysis.tex}
\input{discussion.tex}
\input{conclusions.tex}

  \input{ack.tex}

\bibliographystyle{splncs03}
\bibliography{main}

\begin{appendices}
\renewcommand{\thesection}{\appendixname~\Alph{section}}

\newpage
\section{Additional charts}
\label{sec:charts}
\input{appendixA.tex}
\input{appendixB.tex}
\input{appendixC.tex}
\input{appendixD.tex}

\end{appendices}

\end{document}

%% file: styles.tex
\usepackage[utf8]{inputenc}
\usepackage[english]{babel}

\usepackage{amsmath}
\usepackage{amsfonts}
\usepackage{amssymb}
\usepackage{amsbsy}

\usepackage{color}
\usepackage[usenames,dvipsnames]{xcolor}
\usepackage{mdframed}
\usepackage{tikz}
\usetikzlibrary{decorations.pathmorphing}
\usetikzlibrary{automata,shapes,arrows}
\usepackage{caption,subcaption}

\usepackage{listings}

\usepackage{stmaryrd}
\usepackage{graphicx}
\usepackage{marvosym}

\usepackage{pdfsync}
\usepackage{wasysym}

\usepackage[official]{eurosym} 

\usepackage{fixltx2e} 
\usepackage{xspace} 

\usepackage{nicefrac}

\usepackage{hhline}

\usepackage{multirow}			
\usepackage{pgfplots}			
\usepgfplotslibrary{dateplot} 	

\usepackage[inline,shortlabels]{enumitem} 
\newlist{inlinelist}{enumerate*}{1}
\setlist*[inlinelist,1]{%
	label=(\roman*),
}

\usepackage{xifthen}  

\usepackage{ifthen}

\usepackage[hyphens]{url}
\usepackage{cite}
\usepackage{appendix}

\usepackage[hidelinks]{hyperref}

\hypersetup{
  breaklinks   = true,
  colorlinks   = true, 
  urlcolor     = blue, 
  linkcolor    = blue, 
  citecolor    = red   
}
\hypersetup{final} 

\usepackage{cleveref}

\usepackage[final,nomargin,inline,index]{fixme} 
\fxusetheme{color}

\FXRegisterAuthor{bart}{anbart}{\color{magenta} {\underline{bart}}}
\FXRegisterAuthor{livio}{anlivio}{\color{red} {\underline{livio}}}

\lstdefinelanguage{coco}{
	commentstyle=\color{Gray},
	morecomment=[l]{//},
	morecomment=[s]{/*}{*/},
	classoffset=0,
	morekeywords={honesty,contract,process,system,if,then,else,
	nil,rec,specification,tellAndWait,send,receive,t,tellRetract,tellAndReturn},
	keywordstyle=\color{keyword}\bfseries,
	classoffset=1,
	morekeywords={tell,ask,do,t},
	keywordstyle=[1]\color{ForestGreen},
	classoffset=2,
	morekeywords={unit,int,session,string,boolean},
	keywordstyle=\color{NavyBlue},
	classoffset=0
}

\lstdefinelanguage{maude}{
	classoffset=0,
	morekeywords={mod,ops,eq,endm,red,quit,in},
	keywordstyle=\color{keyword}\bfseries,
	classoffset=1,
	morekeywords={tell,ask,do,t},
	keywordstyle=[1]\color{ForestGreen},
	classoffset=2,
	morekeywords={unit,int,session,string,boolean,exp},
	keywordstyle=\color{NavyBlue},
	classoffset=0
}

\lstdefinelanguage{java}{
	commentstyle=\color{Gray},
	morecomment=[l]{//},
	morecomment=[s]{/*}{*/},
	morestring=[b]",
    stringstyle=\color{NavyBlue},
	classoffset=0,
	morekeywords={public,private,static,final,class,extends,switch,case,break,finally,try,catch,void,int,boolean},
	keywordstyle=\color{keyword}\bfseries
}

%% file: macro.tex
\renewcommand{\vec}[1]{\ensuremath{\textit{\textbf{#1}}}}

\newcommand{\codefont}{\fontsize{10}{10}\selectfont}
\newcommand{\code}[1]{{\tt\codefont {#1}}}

\def\etc{\emph{etc}.\@\xspace}

\newcommand{\eg}{e.g.\@\xspace}
\newcommand{\ie}{i.e.\@\xspace}
\newcommand{\wrt}{w.r.t.\@\xspace}

\newcommand{\Real}[1]{\mathrm{Real}}

\newcommand{\self}{\bullet}

\newcommand{\BTC}{\textup{%
  \leavevmode
  \vtop{\offinterlineskip 
    \setbox0=\hbox{B}%
    \setbox2=\hbox to\wd0{\hfil\hskip-.03em
    \vrule height .3ex width .15ex\hskip .08em
    \vrule height .3ex width .15ex\hfil}
    \vbox{\copy2\box0}\box2}}}

\def\transColor{\color{black}}
\newcommand{\transFmt}[1]{{\transColor{#1}}}

\newcommand{\transT}[1][]{\transFmt{T}_{\transColor{#1}}}
\newcommand{\transTi}[1][]{\transFmt{T'_{\mathit\transColor{#1}}}}

\newcommand{\transIn}[2][]{{\sf in}\ifempty{#1}{}{[{#1}]}: {#2}}
\newcommand{\transInScript}[2][]{{\sf in-script}\ifempty{#1}{}{[{#1}]}: {#2}}
\newcommand{\transOutScript}[3][]{{\sf out-script}\ifempty{#1}{}{[{#1}]}({#2}): {#3}}

\newcommand{\transValue}[2][]{{\sf value}\ifempty{#1}{}{[{#1}]}: {#2}}
\newcommand{\transLockTime}[2][]{{\sf lockTime}\ifempty{#1}{}{[{#1}]}: {#2}}

\newlength\replength
\newcommand\repfrac{.1}

\newcommand\rulewidth{.6pt}
\newcommand\tdashfill[1][\repfrac]{\cleaders\hbox to \replength{%
  \smash{\rule[\arraystretch\ht\strutbox]{\repfrac\replength}{\rulewidth}}}\hfill}
\newcommand\tabdashline{%
  \makebox[0pt][r]{\makebox[\tabcolsep]{\tdashfill\hfil}}\tdashfill\hfil%
  \makebox[0pt][l]{\makebox[\tabcolsep]{\tdashfill\hfil}}%
  \\[-\arraystretch\dimexpr\ht\strutbox+\dp\strutbox\relax]%
}
\newcommand\tdotfill[1][\repfrac]{\cleaders\hbox to \replength{%
  \smash{\raisebox{\arraystretch\dimexpr\ht\strutbox-.1ex\relax}{.}}}\hfill}

\def\cocoColor{\color{MidnightBlue}}
\newcommand{\cocoFmt}[1]{{\cocoColor{\code{#1}}}}

\let\greektau\tau
\renewcommand{\tau}{\cocoFmt{\greektau}}

\newcommand{\ifempty}[3]{%
  \ifthenelse{\isempty{#1}}{#2}{#3}%
}

\newcommand{\compile}[2]{\ifthenelse{\equal{#1}{yes}}{#2}{}}
\newcommand{\hidden}[1]{}

\let\greekgamma\gamma
\def\contrColor{\color{Plum}}
\newcommand{\contrFmt}[1]{{\contrColor{#1}}}

\renewcommand{\gamma}[1][]{\mathord{\contrFmt{\greekgamma}_{\contrFmt{#1}}}}

\crefname{appendix}{appendix}{appendices}
\Crefname{appendix}{Appendix}{Appendices}

\definecolor{LightGrey}{rgb}{0.95,0.95,0.95}
\definecolor{keyword}{HTML}{7F0055}

\newmdenv [linewidth=0pt]{mdNoFramed}
\makeatletter

\newcommand*{\tabminted@finalstrut}[1]{%
  \ifdim\prevdepth>0pt
    \ifdim\dp#1>\prevdepth
      \vskip\dimexpr(\dp#1)-\prevdepth\relax
    \fi
  \else
    \vskip\dimexpr(\dp#1)\relax
  \fi
}
\newcommand*{\@tabmintedend}{%
  \let\@finalstrut\tabminted@finalstrut
}
\makeatother

\newcommand{\opreturn}{OP\_RETURN\xspace}
\newcommand{\pushdata}{PUSH\_DATA\xspace}

\definecolor{rosso}{RGB}{220,57,18}
\definecolor{arancione}{RGB}{255,170,128}
\definecolor{blu}{RGB}{102,140,217}
\definecolor{verde}{RGB}{16,150,24}
\definecolor{viola}{RGB}{153,0,153}
\definecolor{azzurro}{RGB}{173,216,230}

\makeatletter

\tikzstyle{chart}=[
legend label/.style={font={\scriptsize},anchor=west,align=left},
legend box/.style={rectangle, draw, minimum size=5pt},
axis/.style={black,semithick,->},
axis label/.style={anchor=east,font={\tiny}},
]

\tikzstyle{bar chart}=[
chart,
bar width/.code={
		\pgfmathparse{##1/2}
		\global\let\bar@w\pgfmathresult
	},
bar/.style={very thick, draw=white},
bar label/.style={font={\bf\small},anchor=north},
bar value/.style={font={\footnotesize}},
bar width=.75,
]

\tikzstyle{pie chart}=[
chart,
slice/.style={line cap=round, line join=round, very thick,draw=white},
pie title/.style={font={\bf}},
slice type/.style 2 args={
		##1/.style={fill=##2},
		values of ##1/.style={}
	}
]

\pgfdeclarelayer{background}
\pgfdeclarelayer{foreground}
\pgfsetlayers{background,main,foreground}

\newcommand{\pie}[3][]{
		\begin{scope}[#1]
				\pgfmathsetmacro{\curA}{90}
				\pgfmathsetmacro{\r}{1}
				\def\c{(0,0)}
				\node[pie title] at (90:1.3) {#2};
				\foreach \v/\s in{#3}{
						\pgfmathsetmacro{\deltaA}{\v/100*360}
						\pgfmathsetmacro{\nextA}{\curA + \deltaA}
						\pgfmathsetmacro{\midA}{(\curA+\nextA)/2}
						
						\path[slice,\s] \c
						-- +(\curA:\r)
						arc (\curA:\nextA:\r)
						-- cycle;
						\pgfmathsetmacro{\d}{max((\deltaA * -(.5/50) + 1) , .5)}
						
						\begin{pgfonlayer}{foreground}
								\path \c -- node[pos=\d,pie values,values of \s]{$\v\%$} +(\midA:\r);
							\end{pgfonlayer}
						
						\global\let\curA\nextA
					}
			\end{scope}
	}

\newcommand{\legend}[2][]{
		\begin{scope}[#1]
				\path
				\foreach \n/\s in {#2}
				{
						++(1.5,0cm) node[\s,legend box] {} +(0.1cm,0) node[legend label] {\n}
					}
				;
		\end{scope}
}

%% file: abstract.tex
\begin{abstract}
  The Bitcoin protocol allows to save arbitrary data on the blockchain
  through a special instruction of the scripting language,
  called \opreturn. 
  A growing number of protocols exploit this feature to 
  extend the range of applications of the Bitcoin blockchain
  beyond transfer of currency.
  A point of debate in the Bitcoin community is whether 
  loading data through \opreturn 
  can negatively affect the performance of the Bitcoin network
  with respect to its primary goal.
  This paper is an empirical study of the usage of \opreturn
  over the years.
  We identify several protocols based on \opreturn,
  which we classify by their application domain.
  We measure the evolution in time of the usage of each protocol,
  the distribution of \opreturn transactions by application domain,
  and their space consumption.
\end{abstract}

%% file: intro.tex
\section{Introduction}

Bitcoin was the first decentralized digital currency
to be created, and now it is the most widely used, 
with a market capitalization of \mbox{$\sim 20$} billions USD%
\footnote{Source: \href{http://coinmarketcap.com}{coinmarketcap.com}, accessed on February 28th, 2017.}.
Technically, the Bitcoin network is a peer to peer system, 
where users can securely transfer currency without 
the intermediation of a trusted authority. 
Transactions of currency are gathered in blocks, 
that are added to a public data structure called \textit{blockchain}. 
The consensus algorithm of Bitcoin guarantees that, 
for an attacker to be able to alter an existing block, 
she must control the majority of the computational resources of the network~\cite{Garay15eurocrypt}. 
Hence, attacks aiming at incrementing one's balance,
\eg by deleting transactions that certify payments to other users,
are infeasible in practice. 
This security property is often rephrased by saying that 
the blockchain can be seen as an \emph{immutable} data structure.

Although the main goal of Bitcoin is to transfer digital currency, 
the immutability and openness of its blockchain have inspired the 
development of new protocols, 
which ``piggy-back'' metadata on transactions
in order to implement a variety of applications beyond cryptocurrency.
For instance, some protocols allow to certify the existence of a document 
(\eg,~\cite{StamperyWebsite, FactomWebsite, ProofOfExistenceWebsite}), 
while some others allow to track the ownership of a digital or a physical asset 
(\eg,~\cite{OpenAssetsWebsite, ColuWebsite, OmniWebsite}).
Many of these protocols save metadata on the blockchain 
by using an instruction called \opreturn, which is part of the Bitcoin scripting language.

A debate about the scalability of Bitcoin has been taking place 
over the last few years~\cite{BicoinScalabilityWiki,BitcoinScalabilityDebateCryptocoinsnews,BitcoinScalabilityDebateCoindesk}. 
In particular, users argue over whether the blockchain should allow for storing spurious data, not inherent to currency transfers.
Although many recent works analyse the Bitcoin 
blockchain~\cite{moser2015trends,baqerstressing,ron2013quantitative, reid2013analysis}, 
as well as some services related to 
\opreturn~\cite{BitcoinOpReturnDescription,SmartbitOpReturnStatistics,OpReturnWebsiteStatistics,KaikoWebsiteStatistics}, 
many relevant questions are still open.
What is the impact of the data attached to \opreturn  
on the size of the blockchain? 
Which kinds of blockchain-based applications are exploiting the \opreturn instruction, and how?

\paragraph{Contributions.}
We analyse the usage of \opreturn throughout the Bitcoin blockchain, 
collecting a total of 1,887,708 \opreturn transactions. 
We investigate to which protocols \opreturn transactions belong, 
identifying 22 distinct protocols (associated to 51\% of these transactions). 
We find that 15\% of this total are \textit{empty} transactions, 
which attach no metadata to \opreturn.
By studing the usage of \opreturn over time, 
we identify several transaction peaks related to empty transactions, 
and we show that they are mainly caused by stress tests 
and spam attacks happened in summer 2015. 
We classify protocols according to their application domain, 
and we study the numerical proportion of these applications.
Finally, we measure the size of \opreturn metadata, 
and the proportion between the size of \opreturn transactions
and the overall size of the transactions in the blockchain.
To the best of our knowledge, ours is the widest investigation 
about the usage of \opreturn.
All our analyses are supported by a tool we have developed.
The sources of our tool, as well as the experimental data, are available at~\cite{BitcoinOpReturnExplorer}.

%% file: background.tex
\section{Background on Bitcoin}
\label{sec:background}

Bitcoin~\cite{bitcoin} is a decentralized infrastructure to exchange virtual currency
--- the \emph{bitcoins}.
The transfers of currency, called \emph{transactions}, 
are the basic elements of the system.
The transactions are recorded on a public, append-only data structure,
called \emph{blockchain}.
To illustrate how Bitcoin works,
we consider two transactions $\transT[0]$ and $\transT[1]$ 
of the following form:
\begin{center}
  \begin{tabular}{|l|}
    \hhline{-}
    \multicolumn{1}{|c|}{$\transT[0]$} \\
    \hhline{|-|}
    \transIn{$\cdots$} \\
    \transInScript{$\cdots$} \\
    \hhline{|-|}
    \transOutScript{$\transT,\sigma$}{${\it ver}_k(\transT,\sigma)$} \\
    \transValue{$v_0$} \\
    \hhline{|-|}
  \end{tabular}
  \qquad\qquad
  \begin{tabular}{|l|}
    \hhline{-}
    \multicolumn{1}{|c|}{$\transT[1]$} \\
    \hhline{|-|}
    \transIn{$\transT[0]$} \\
    \transInScript{${\it sig}_k(\self)$} \\
    \hhline{|-|}
    \transOutScript{$\cdots$}{$\cdots$} \\
    \transValue{$v_1$} \\
    \hhline{|-|}
  \end{tabular}
\end{center}

The transaction $\transT[0]$ contains a value $v_0$ bitcoins.
Anyone can \emph{redeem} the amount of bitcoins in $\transT[0]$ 
by putting on the blockchain a transaction (\eg, $\transT[1]$),
whose {\sf in} field contains the identifier of $\transT[0]$ 
(the hash of the whole transaction, displayed as $\transT[0]$ in the figure)
and whose {\sf in-script} contains values making the {\sf out-script}%
\footnote{
  {\sf in-script}/{\sf out-script} are called {\sf scriptPubKey}/{\sf scriptSig} in the
  Bitcoin wiki. 
}
of $\transT[0]$, 
a programmable boolean function, evaluate to true. 
When this happens, 
the value of $\transT[0]$ is transferred to the new transaction $\transT[1]$,
and $\transT[0]$ becomes unredeemable.
A subsequent transaction can then redeem $\transT[1]$ likewise.

In the transaction $\transT[0]$ above, 
the {\sf out-script} just checks the digital signature $\sigma$
on the redeeming transaction $\transT$ \wrt a given key~$k$.
We denote with ${\it ver}_k(\transT,\sigma)$ the signature verification,
and with ${\it sig}_k(\self)$ 
the signature of the enclosing transaction ($\transT[1]$ in our example),
including all the parts of the transaction but its {\sf in-script}
(obviously, because it contains the signature itself).

Now, assume that $\transT[0]$ is redeemable on the blockchain 
when someone tries to append~$\transT[1]$.
The Bitcoin network accepts the redeem if 
\begin{inlinelist}
\item $v_1 \leq v_0$, and
\item the {\sf out-script} of $\transT[0]$, 
  applied to to $\transT[1]$ and to the signature ${\it sig}_k(\self)$, evaluates true.
\end{inlinelist}

The previous example is a special case of a Bitcoin transaction:
the general form is displayed in~\Cref{fig:transaction-template}.
First, there can be multiple inputs and outputs (denoted with array notation in the figure),
and each output has its own \mbox{{\sf out-script}} and value.
Since each output can be redeemed independently, 
{\sf in} fields must specify which one they are redeeming
($\transT[0]\transColor{[n_0]}$ in the figure).
A transaction with multiple inputs redeems \emph{all} the 
(outputs of) transactions in its {\sf in} fields, 
providing a suitable {\sf in-script} for each of them.
To be valid, the sum of the values of all the inputs must be greater or equal to 
the sum of the values of all outputs.
The \emph{Unspent Transaction Output} (in short, UTXO) 
is the set of redeemable outputs of all transactions included in the blockchain. 
To be valid, a transaction must only use elements of the UTXO as inputs.

In its general form, the {\sf out-script}
is a program in a non Turing-complete scripting language,
which features a limited set of logic, arithmetic, and cryptographic operators.
The {\sf lockTime} field specifies the earliest moment in time when the transaction can appear on the blockchain.

\begin{figure}[t]
  \centering
  \scalebox{0.9}{
  \begin{tabular}{cc}
    \begin{subfigure}[b]{.4\textwidth}
      \begin{tikzpicture}[scale=1, transform shape, >=latex]
        \node at (1,-1) {
          \begin{tabular}{|l|}
            \hhline{-}
            \multicolumn{1}{|c|}{$\transT$} \\
            \hhline{|-|}
            \transIn[0]{$\transT[0]\transColor{[n_0]}$} \\
            \transInScript[0]{$\cdots$} \\
            \tabdashline
            \multicolumn{1}{|c|}{\vdots} \\
            \hhline{|-|}
            \transOutScript[0]{$\transTi[0],\vec{w}_0$}{$\cdots$\qquad} \\
            \transValue[0]{$v_0$} \\
            \tabdashline
            \multicolumn{1}{|c|}{\vdots} \\
            \hhline{|-|}
            \transLockTime{$s$} \\
            \hhline{|-|}
          \end{tabular}
        };
      \end{tikzpicture}
      \caption{General form of transactions.}
      \label{fig:transaction-template}
    \end{subfigure}
    \hspace{0pt}
 &
   \begin{subfigure}[b]{.5\textwidth}
     \begin{tikzpicture}[scale=1, transform shape, >=latex]
       \node at (1,-1) {
         \begin{tabular}{|l|}
           \hhline{-}
           \multicolumn{1}{|c|}{$\transT$} \\
           \hhline{|-|}
           \transIn[0]{...} \\
           \transInScript[0]{...} \\
           \tabdashline
           \multicolumn{1}{|c|}{\vdots} \\
           \hhline{|-|}
           \transOutScript[0]{...}{\opreturn ``$EW Hello!$''} \\
           \transValue[0]{$0$} \\
           \tabdashline
           \multicolumn{1}{|c|}{\vdots} \\
           \hhline{|-|}
         \end{tabular}
       };
     \end{tikzpicture}
     \caption{An \opreturn transaction.}
     \label{fig:opreturnexample}
   \end{subfigure}
  \vspace{-20pt}
  \end{tabular}
}
\end{figure}

\paragraph{Writing metadata in transactions.}

Bitcoin transactions do not provide a field where one can save arbitrary data. 
Nevertheless, users have devised various creative ways to encode data in transactions.
A first method is to abuse the standard \emph{Pay-to-PubkeyHash} script%
\footnote{\href{https://en.bitcoin.it/wiki/Transaction\#Pay-to-PubkeyHash}{en.bitcoin.it/wiki/Transaction\#Pay-to-PubkeyHash}},
which implements the signature verification ${\it ver}_k$ seen before
(actually, the script does not contain the public key $k$, but its hash $h = H(k)$).
To make the script evaluate to true, the redeeming transaction $\transT$ 
has to provide the signature $\sigma$ and a public key $k$
such that $H(k) = h$ and ${\it ver}_k(\transT,\sigma)$.
One can store an arbitrary message $m$ (a few bytes long) within the {\sf out-script},
by writing $m$ in place of the hash $h$. 
Since computing a value $k$ such that $H(k) = m$ (\ie, a preimage of $m$) 
and a signature $\sigma$ such that ${\it ver}_k(\transT,\sigma)$
are computationally hard operations, outputs crafted in this way are unspendable in practice. 
However, these outputs are not easily distinguishable from the spendable ones,
hence the nodes of the Bitcoin network must keep them in their UTXO set~\cite{CoindeskBitcoinDevCore5}. 
Since this set is usually stored in RAM for efficiency concerns~\cite{DelayedTxoCommitments},
this practice negatively affects the memory consumption of nodes~\cite{baqerstressing}.

The \opreturn instruction allows to save metadata on the blockchain, 
as shown in~\Cref{fig:opreturnexample}%
\footnote{Hash: \href{https://blockchain.info/tx/d84f8cf06829c7202038731e5444411adc63a6d4cbf8d4361b86698abad3a68a}{d84f8cf06829c7202038731e5444411adc63a6d4cbf8d4361b86698abad3a68a}}.
However, unlike \emph{Pay-to-PubkeyHash},
an {\sf out-script} containing \opreturn always evaluates to false,
hence the output is provably unspendable, and its transaction can be safely removed from the UTXO.
In this way, \opreturn overcomes the UTXO consumption issue highlighted above.
Although the \opreturn instruction has been part of the scripting language 
since the first releases of Bitcoin, 
originally it was considered \emph{non-standard} by nodes, so 
transactions containing \opreturn were difficult to reliably get mined.
In March 2014~\cite{Bitcoin0.9.0}, \opreturn became standard,
meaning that all nodes started to relay unconfirmed \opreturn transactions%
\footnote{
  Regarding the use of \opreturn, 
  the release notes of Bitcoin Core version 0.9.0 state that:
  \emph{``This change is not an endorsement of storing data in the blockchain.''}
  At the same time, some Bitcoin explorers,
  (\eg~\href{https://blockchain.info/}{blockchain.info}, 
  \href{https://blockexplorer.com/}{blockexplorer.com}, 
  \href{https://www.smartbit.com.au/}{smartbit.com}) 
  allow to inspect data encoded in \opreturn scripts.
}. 
The limit for storing data in an \opreturn
was originally planned to be 80 bytes, 
but the first official client supporting the instruction, \ie the release 0.9.0~\cite{Bitcoin0.9.0}, 
allowed only 40 bytes.
This animated a long debate~\cite{BitcoinOpReturnBattle, BitcoinPull5075, BitcoinPull5286, CounterpartyOpenLetter}. 
From the release 0.10.0~\cite{Bitcoin0.10.0} 
nodes could choose whether to accept or not \opreturn transactions, and set a maximum for their size. 
The release 0.11.0~\cite{Bitcoin0.11.0} extended the data limit to 80 bytes,
and the release 0.12.0~\cite{Bitcoin0.12.0} to a maximum of 83 bytes. 

%% file: methodology.tex
\section{Methodology for classifying \opreturn transactions}
\label{sec:methodology}

We discuss our methodology for
identifying protocols that use \opreturn.

We gather all the \opreturn transactions 
from the origin block up to the block number 453,200 
(added on 2017/02/15).
We end up with a set of 1,887,708 \opreturn transactions. 
For each of them, we save the following data in a database:
\begin{inlinelist}
\item the hash of the transaction;
\item the hash of the enclosing block;
\item the timestamp of the block;
\item the metadata attached to the \opreturn.
\end{inlinelist}

Next, we detect to which protocols the \opreturn transactions belong.
Usually, a protocol is identified by the first few bytes of metadata
attached to the \opreturn,
but the exact number of bytes may vary from protocol to protocol.
Hence, we associate \opreturn transactions to protocols as follows:
\begin{enumerate}
\item we search the web for known associations 
between identifiers and protocols;
\item we accordingly classify the \opreturn transactions that begin with one of the identifiers obtained at step 1;
\item on the remaining \textit{unknown transactions}, 
we perform a frequency analysis of the first few bytes of metadata, 
to discover new protocol identifiers.
\end{enumerate}

In more details, in the first step we query Google
to obtain public identifier/protocol bindings. 
For instance, the query \mbox{\textit{``Bitcoin \opreturn''}}, 
returns $\sim$26,500 results, and we manually inspect
the first few pages of them.
Note that a protocol can be associated with more than one identifier
(\eg, Stampery, Blockstore~\cite{ali2016blockstack}, 
Remembr, CryptoCopyright),
or even do not have any identifier. 
In this way we obtain 22 protocols associated to 33 identifiers;
further, we find 3 protocols that do not use any identifier
(Counterparty, Diploma~\cite{DiplomaWebsite}, Chainpoint~\cite{ChainpointWebsite}).

The second step is performed by our tool: 
it associates 970,374 transactions to a protocol
($\sim$51\% of the total \opreturn transactions).
The other transactions are classified either 
as \textit{empty} or \textit{unknown}. 
Empty transactions have no data attached to the \opreturn instruction
(296,491 transactions, $\sim$15\% of the total); 
unknown transactions have no known identifier 
(620,843 transactions, $\sim$32\% of the total).

The final step analyses unknown transactions,
attempting to discover new protocol identifiers.
Since identifiers may have different lengths, 
we gather the first $D$ bytes of unknown transactions,
for $D$ ranging from 1 to 12,
and we perform a frequency analysis of these strings.
This analysis does not reveal relevant statistical anomalies
(roughly, the strings are uniformly distributed),
hence this step does not yield any new identifier.
\Cref{alg:recurrent} details this search, 
which is executed with the following parameters:
${\it D} = 12$, $\delta = 2$, $N = 100$. 

\begin{algorithm}[t]
  \scalebox{0.9}{
  \begin{minipage}{1\linewidth}
  \caption{Detect protocol identifiers}
  \label{alg:recurrent}
	\begin{algorithmic}
		\State unknownTx $\leftarrow$ \textit{set of all unknown transactions}
                \State Codes $\leftarrow \emptyset$
		\For{i $\leftarrow$ 1 to $D$}
			\State H $\leftarrow$ \textit{new hash table from protocol identifiers to number of occurrences}
			\ForAll{tx $\in$ unknownTx}
				\State	code $\leftarrow$ tx.substring(i)
                                \Comment first i characters of tx
                                \If {(H.contains(code))}
                                \State H.code $\leftarrow$ H(code)+1
                                \textbf{else} H.code $\leftarrow$ 1
                                \EndIf
			\EndFor
			\State expectedOccurrences $\leftarrow$ unknownTx.size() / pow(16,i)
			\ForAll{h $\in$ H}
				\If{(h.occurrences $>$ expectedOccurrences * $\delta$ \textbf{and} h.occurrences $>$ $N$) } 
                                \State Codes $\leftarrow$ Codes $\cup$ \{h.code\}
				\EndIf
			\EndFor
		\EndFor
                \State \textbf{return} Codes
	\end{algorithmic}
\end{minipage}
}
\end{algorithm}

%% file: qualitative.tex
\section{Qualitative analysis of \opreturn transactions}
\label{sec:qualitative}

We now classify the protocols obtained in~\Cref{sec:methodology},
associating each protocol to a \emph{category}
that describes its intended application domain. 
To this purpose, we manually inspect the web pages of each protocol.

\begin{description}
\item[Assets] gathers protocols that exploit the immutability of the blockchain 
  to certify ownership, exchange, and eventually the value of real-world assets. 
  Metadata in transactions are used to specify \eg
  the value of the asset, 
  the amount of the asset transferred, the new owner, \etc
  
\item[Document notary] includes protocols for certifying 
  the ownership and timestamp of a document.
  A user can publish the hash of a document in a transaction,
  and in this way he can prove its existence and integrity.
  Similarly, signatures can be used to certify ownership.
  
\item[Digital arts] includes protocols for 
  declaring access right and copy rights 
  on digital arts files, like \eg photos or music. 
  
\item[Other] includes protocols whose goals differ from the ones above. 
  For instance, \textit{Eternity Wall}\cite{EternityWallWebsite} 
  allows users to store short text messages on the blockchain;
  \textit{Blockstore}~\cite{BlockstoreWebsite} is a generic 
  key-value store, on top of which more complex protocols 
  can be implemented%
  \footnote{Hereafter we aggregate all the protocols built upon \textit{Blockstore}, by identifying them with \textit{Blockstore} itself.}.

\item[Empty] includes protocols that do not attach any data to \opreturn.
  
\item[Unknown] includes  protocols for which we have not been able to detect
  an identifier (possibly, because they do not use any).
\end{description}

\noindent
We report our classification of
protocols in the first two columns of~\Cref{fig:table}.
Due to the \opreturn space limit, 
long pieces of metadata
require to be split in many transactions, and higher fees.
Hence, \textit{assets} protocols usually feature complex rules, 
have space-efficient representations of data,
and often propose off-chain solutions~\cite{ColuTorrents}.
We distinguish document notary protocols from digital arts protocols for the following reason. 
Most document notary applications do not require users to provide their documents to the application,
and the main purpose of the protocol (certifying ownership) 
can be fulfilled also when the application is no longer live.
Instead, digital arts application usually need to gather user documents,
and require interactions with users, \eg they often play the role of broker between producers and consumers.

%% file: analysis.tex
\section{Quantitative analysis of \opreturn transactions}

\Cref{fig:table} shows some statistics about \opreturn transactions.
The first column indicates the protocol categories, introduced in~\Cref{sec:qualitative}. 
The second and third columns show, respectively, the protocol names and the associated identifiers. 
The fourth column shows the date in which the protocol generated the first transaction. 
Since transactions do not have a ``date'' field, we infer dates from the timestamp of the block containing the transaction.
The next two columns count the total number of transactions,
and the total size (in bytes) of the \opreturn data contained therein. 
To compute the size we only consider the metadata, 
\ie we do not count neither the \opreturn instruction nor the other fields of the transaction. 
The last column shows the average size of the transaction metadata.

\begin{table}[t]
  \begin{center}
    \small
    \scalebox{0.75}{
      \begin{tabular}{|c|c|c|c|c|c|c|}
        \hline
        \textbf{Category} & \textbf{Protocol} & \textbf{Identifiers} & \textbf{First trans.} & \textbf{Tot.\ trans.} & \textbf{Tot.\ Size} & \textbf{Avg.\ Size} \\
        \hline\hline
        \multirow{6}{3em}{\textbf{Assets}} & \href{https://www.colu.com/}{Colu} & CC & 2015/07/09 & 237,479 & 4,290,388 & 18.0\\
                 & \href{http://coinspark.org/}{CoinSpark} & SPK & 2014/07/02 & 28,026 & 956,904 & 34.1\\ 
                 & \href{https://github.com/OpenAssets}{OpenAssets} & OA & 2014/05/03 & 133,570 & 1,728,350 & 12.9\\
                 & \href{http://www.omnilayer.org/}{Omni} & omni & 2015/08/10 & 105,979 & 2,132,565 & 20.1\\
                 & \href{http://counterparty.io/}{Counterparty} & N/A & N/A & N/A & N/A & N/A\\
                 & \textbf{Total} & \textbf{-} & \textbf{-} & \textbf{505,054} & \textbf{9,108,207} & \textbf{18.0}\\
        \hline \hline
        \multirow{15}{5.2em}{\textbf{Document Notary}} & \href{https://www.factom.com/}{Factom} & Factom!!, FACTOM00, Fa, FA & 2014/04/11 & 74,159 & 2,966,234 & 40.0\\
                 & \href{https://stampery.com/}{Stampery} & S1, S2, S3, S4, S5 & 2015/03/09 & 74,249 & 2,627,540 & 35.4\\
                 & \href{https://proofofexistence.com/}{Proof of Existence} & DOCPROOF & 2014/04/21 & 5,262 & 210,433 & 40.0\\
                 & \href{https://blocksign.com/}{Blocksign} & BS & 2014/08/04 & 1,460 & 55,192 & 37.8\\
                 & \href{https://crypto-copyright.com/}{CryptoCopyright} & CryptoTests-, CryptoProof- & 2014/08/02 & 46 & 1,840 & 40\\
                 & \href{https://stampd.io/}{Stampd} & STAMPD\#\# & 2015/01/03 & 473 & 18,867 & 39.9\\
                 & \href{https://bitproof.io/}{BitProof} & BITPROOF & 2015/02/25 & 758 & 30,320 & 40\\
                 & \href{https://github.com/thereal1024/ProveBit}{ProveBit} & ProveBit & 2015/04/05 & 57 & 2,280 & 40\\
                 & \href{https://remembr.io/}{Remembr} & RMBd, RMBe & 2015/08/25 & 28 & 1,128 & 40.3\\
                 & \href{https://originalmy.com/}{OriginalMy} & ORIGMY & 2015/07/12 & 126 & 4,788 & 38\\
                 & \href{http://lapreuve.eu/explication.html}{LaPreuve} & LaPreuve & 2014/12/07 & 67 & 2,623 & 39.1\\
                 & \href{http://digitalcurrency.unic.ac.cy/free-introductory-mooc/academic-certificates-on-the-blockchain/}{Nicosia} & UNicDC & 2014/09/12 & 20 & 684 & 34.2\\
                 & \href{http://www.chainpoint.org/}{Chainpoint} & N/A & N/A & N/A & N/A & N/A\\
                 & \href{http://diploma.report/}{Diploma} & N/A & N/A & N/A & N/A & N/A\\
                 & \textbf{Total} & \textbf{-} & \textbf{-} & \textbf{156,705} & \textbf{5,921,929} & \textbf{37.8}\\
        \hline \hline
        \multirow{4}{3em}{\textbf{Digital Arts}} & \href{https://monegraph.com/}{Monegraph} & MG & 2015/06/28 & 63,278 & 2,317,151 & 36.6\\
                 & \href{https://blockai.com/}{Blockai} & 0x1f00 & 2015/01/09 & 527 & 34,225 & 64.9\\
                 & \href{https://www.ascribe.io}{Ascribe} & ASCRIBE & 2014/12/19 & 40,859 & 847,641 & 20.7\\
                 & \textbf{Total} & \textbf{-} & \textbf{-} & \textbf{104,664} & \textbf{3,199,017} & \textbf{30.6}\\
        \hline \hline
        \multirow{5}{3em}{\textbf{Other}} & \href{https://eternitywall.it/}{Eternity Wall} & EW & 2015/06/24 & 3,715 & 160,191 & 43.1\\
                 & \href{https://github.com/blockstack/blockchain-id/wiki/Blockstore}{Blockstore} & id, 0x5888, 0x5808 & 2014/12/10 & 191,907 & 5,494,174 & 28.6\\
                 & \href{https://www.smartbit.com.au/}{SmartBit} & SB.D & 2015/11/24 & 8,329 & 299,844 & 36\\
                 & \textbf{Total} & \textbf{-} & \textbf{-} & \textbf{203,951} & \textbf{5,954,209} & \textbf{29.2}\\
        \hline \hline
        \multirow{1}{3em}{\textbf{Empty}} & {\bf Total} & \textbf{-} & 2014/03/20 & {\bf 296,491} & {\bf 0} & {\bf 0} \\
        \hline \hline
        \multirow{1}{4.5em}{\textbf{Unknown}} & {\bf Total} & \textbf{-} & 2014/03/12 & {\bf 620,843} & {\bf 20,023,345} & {\bf 32.3} \\
        \hline \hline
        \textbf{TOTAL} & \textbf{-} & \textbf{-} & 2014/03/12 & \textbf{1,887,708} & \textbf{44,206,707} & \textbf{23.4} \\
        \hline
      \end{tabular}
    }
  \end{center}
  \caption{Statistics about \opreturn protocols.}
  \label{fig:table}
\end{table}

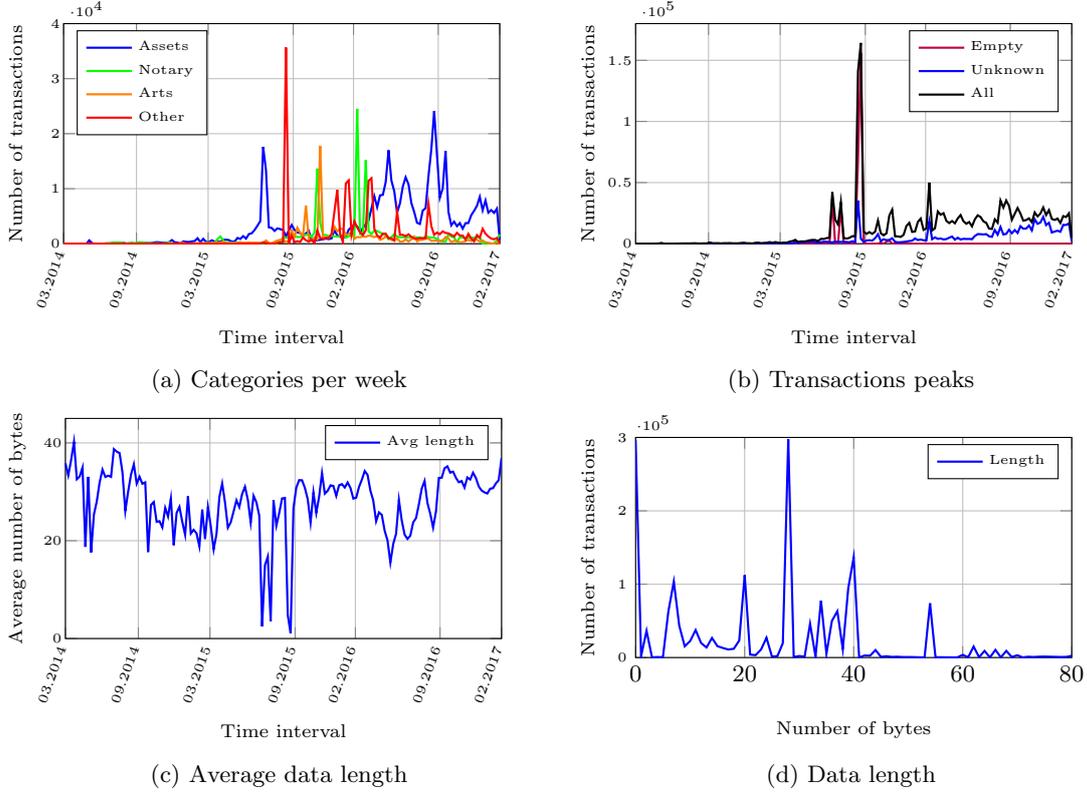
\begin{figure}[t]
  \hspace{-50pt}
  \begin{tabular}{cc}
    \begin{subfigure}[b]{.6\textwidth}
			\begin{tikzpicture}
			\begin{axis}[
			width  = 1\linewidth,
			height = 4.5cm,
			date coordinates in=x, date ZERO=2014-03-12,
			xmin=2014-03-12, xmax=2017-02-15,
			xtick={2014-03-12, 2014-09-07, 2015-03-02, 2015-09-27, 2016-02-22, 2016-09-17, 2017-02-15},
			ymin=0, ymax=40000,
			xmajorgrids = true,
			ymajorgrids = true,
			xticklabel style={rotate=70, anchor=east, xticklabel}, xticklabel=\tiny\month.\year,
			yticklabel style={font=\tiny,xshift=0.5ex},
			legend pos=north west, legend style={font=\tiny}, legend cell align=left,
			xlabel absolute, xlabel style={yshift=-0.5cm}, xlabel={\scriptsize Time interval},
			ylabel absolute, ylabel style={yshift=-0.6cm}, ylabel={\scriptsize Number of transactions}
			]
			\pgfplotstableread[col sep=comma]{results/Categories.csv}\data
			\addplot [color=blue, thick] table[x index = {0}, y index = {1}]{\data};
			\addplot [color=green, thick] table[x index = {0}, y index = {2}]{\data};
			\addplot [color=orange, thick] table[x index = {0}, y index = {3}]{\data};
			\addplot [color=red, thick] table[x index = {0}, y index = {4}]{\data};
			\legend{Assets, Notary, Arts, Other}
			\end{axis}
			\end{tikzpicture}
      \caption{Categories per week}
      \label{fig:Groups}
    \end{subfigure}
    \hspace{0pt}
    &
      \begin{subfigure}[b]{.6\textwidth}
			\begin{tikzpicture}
			\begin{axis}[
			width  = 1\linewidth,
			height = 4.5cm,
			date coordinates in=x, date ZERO=2014-03-12,
			xmin=2014-03-12, xmax=2017-02-15,
			xtick={2014-03-12, 2014-09-07, 2015-03-02, 2015-09-27, 2016-02-22, 2016-09-17, 2017-02-15},
			ymin=0, ymax=180000,
			xmajorgrids = true,
			ymajorgrids = true,
			xticklabel style={rotate=70, anchor=east, xticklabel}, xticklabel=\tiny\month.\year,
			yticklabel style={font=\tiny,xshift=0.5ex},
			legend pos=north east, legend style={font=\tiny}, legend cell align=left,
			xlabel absolute, xlabel style={yshift=-0.5cm}, xlabel={\scriptsize Time interval},
			ylabel absolute, ylabel style={yshift=-0.6cm}, ylabel={\scriptsize Number of transactions}
			]
			\pgfplotstableread[col sep=comma]{results/Peaks.csv}\data
			\addplot [color=purple, thick] table[x index = {0}, y index = {1}]{\data};
			\addplot [color=blue, thick] table[x index = {0}, y index = {2}]{\data};
			\addplot [color=black, thick] table[x index = {0}, y index = {3}]{\data};
			\legend{Empty, Unknown, All}
			\end{axis}
			\end{tikzpicture}
        \caption{Transactions peaks}
        \label{fig:Peaks}
      \end{subfigure}
    \\
    \begin{subfigure}[b]{.6\textwidth}
			\begin{tikzpicture}
			\begin{axis}[
			width  = 1\linewidth,
			height = 4.5cm,
			date coordinates in=x, date ZERO=2014-03-12,
			xmin=2014-03-12, xmax=2017-02-15,
			xtick={2014-03-12, 2014-09-07, 2015-03-02, 2015-09-27, 2016-02-22, 2016-09-17, 2017-02-15},
			ymin=0, ymax=45,
			xmajorgrids = true,
			ymajorgrids = true,
			xticklabel style={rotate=70, anchor=east, xticklabel}, xticklabel=\tiny\month.\year,
			yticklabel style={font=\tiny,xshift=0.5ex},
			legend pos=north east, legend style={font=\tiny}, legend cell align=left,
			xlabel absolute, xlabel style={yshift=-0.5cm}, xlabel={\scriptsize Time interval},
			ylabel absolute, ylabel style={yshift=-0.6cm}, ylabel={\scriptsize Average number of bytes}
			]
			\pgfplotstableread[col sep=comma]{results/LengthByTime.csv}\data
			\addplot [color=blue, thick] table[x index = {0}, y index = {1}]{\data};
			\legend{Avg length}
			\end{axis}
			\end{tikzpicture}
      \caption{Average data length}
      \label{fig:AverageLength}
    \end{subfigure}
    \hspace{0pt}
    &
      \begin{subfigure}[b]{.6\textwidth}
			\begin{tikzpicture}
			\begin{axis}[
			width  = 1\linewidth,
			height = 4.5cm,
			xmin=0, xmax=80,
			ymin=0, ymax=300000,
			xmajorgrids = true,
			ymajorgrids = true,
			yticklabel style={font=\tiny},
			legend pos=north east, legend style={font=\tiny}, legend cell align=left,
			xlabel absolute, xlabel style={yshift=-0.2cm}, xlabel={\scriptsize Number of bytes},
			ylabel absolute, ylabel style={yshift=-0.6cm}, ylabel={\scriptsize Number of transactions}
			]
			\pgfplotstableread[col sep=comma]{results/SizeDistribution.csv}\data
			\addplot [color=blue, thick] table[x index = {0}, y index = {1}]{\data};
			\legend{Length}
			\end{axis}
			\end{tikzpicture}
        \caption{Data length}
        \label{fig:Length}
      \end{subfigure}
  \end{tabular}
  \caption{Usage and size of \opreturn transactions.}
  \label{fig:TransactionsAndLength}
\end{figure}

%% file: discussion.tex
\subsection{Overall statistics}
\label{subsec:ResultGeneral}
We detect 1,887,708 \opreturn transactions, distributed into 98,233 blocks, 
by scanning the blockchain until block number 453,200.
Overall, \opreturn transactions constitute $\sim 0.96\%$ 
of the total transactions in the blockchain, 
and $\sim 1.16\%$ of the portion of the blockchain from 2014/03/12 
(when the first \opreturn transaction appeared).
Although the former measurement considers 7 years of transactions while the latter only considers 
the last 3 years, we note that the values are very close. 
We explain this fact by observing that the daily number of transactions rapidly increased since July 2014.

\subsection{Transaction peaks}
\label{subsec:ResultPeaks}
\Cref{fig:Groups,fig:Peaks} display the number of \opreturn transactions per week, 
from 2014/03 (date of the first \opreturn transaction) to 2017/02 (end of our extraction). 
In the graph we note several peaks, that we explain as follows:
\begin{enumerate}

\item $\sim$100,000 transactions from 2015/07/08 to 2015/08/05.
This peak is mainly composed of two different peaks of \textit{empty} transactions:
the july peak ($\sim$36,900 transactions from 2015/07/08 to 2015/07/10) and the
august peak ($\sim$29,200 transactions from 2015-08-01 to 2015-08-03). 
Both peaks seem to be caused by a spam campaign that resulted in a DoS attack on Bitcoin
which happened in the same period, as reported in~\cite{baqerstressing}.

\item $\sim$300,000 transactions from 2015/09/09 to 2015/09/23.
This second peak is the highest and longest-lasting one. 
As before, it is mainly caused by \textit{empty} transactions ($\sim$223,000),
although here we also observe a component of \textit{unknown} and \textit{blockstore} 
transactions ($\sim$35,000 each). 
The work~\cite{baqerstressing} detects a spike also in this period, precisely around 2015/09/13,
where an anonymous group performed a stress-test on the network with a \textit{money drop}. 
This involves a public release of private keys, with the 
aim to cause a big race which would cause a large number of \textit{double-spend} transactions.

\item $\sim$50,000 transactions from 2016/03/02 to 2016/03/09.
The last peak is due to the sum of two different peaks: 
\textit{unknown} (about 18,000) and \textit{stampery} (about 23,000) transactions.
We conjecture that this peak is caused by the testing and bootstrap of protocols.

\end{enumerate}

We observe that the Bitcon blockchain has also other peaks, not related to \opreturn transactions.
For instance, starting from the 2015/05/22 and for a duration of 100 blocks,
the Bitcoin network was targeted by a stress test~\cite{BitcoinStressTest},
during which the network was flooded with a huge number of transactions.
Actually, the usage of \opreturn transactions in the period of this peak 
does not seem to diverge from their normal usage.

\subsection{Space consumption}
\label{subsec:ResultSize}

A debated topic in the Bitcoin community is whether it is acceptable or not 
to save arbitrary data in the blockchain. 
The sixth column in~\Cref{fig:table} shows, 
for each protocol, the total size of metadata
(\ie, not considering the bytes of the instructions \opreturn and \pushdata). 
The last row of~\Cref{fig:table} 
shows that the total size of metadata is \mbox{$\sim$ 42} MB 
(in the same date, the size of the whole blockchain is \mbox{$\sim$ 102} GB). 
\Cref{fig:AverageLength} shows the average length of the data for each week. 

Generally, the average length of metadata is less than 40 bytes, despite the 
extension to 80 bytes introduced on 2015/07/12. 
Peaks down on the same period are related to the \textit{empty} transactions discussed in~\Cref{subsec:ResultPeaks}. 
\Cref{fig:Length} represents the number of transactions with a given data length: also this chart confirms a
small number of transactions that use more than the half of the available space. 
Note that the discussed peak appears also in this chart, in correspondence of the 0 value. 
From the last column of~\Cref{fig:table} we see that only the size of \textit{Blockai} metadata is close to 80 bytes. 
Several \textit{document notary} protocols take 40 bytes on average:
this depends from their identifiers, composed of 16 bytes, and from the size of the hash they save.
Generally, \textit{document notary} protocols carry longer data than the other protocols. 

We now evaluate the minimum space consumption of the \opreturn transactions on the whole blockchain.
First, we observe that an \textit{empty} transaction with one input and one output has a total size of 156 bytes. 
From~\Cref{fig:table} we see that \opreturn transaction carry $\sim$23.4 bytes of metadata, on average.
Hence, we approximate the average size of \opreturn transaction as $\sim$179.4 bytes,
and so an approximation of the space consumption of all the \opreturn transactions is \mbox{$\sim$323 MB}. 

Finally, we estimate the ratio between the total size of \opreturn transactions and the size of all the transactions on the blockchain.
The block header has size 97 bytes at most. 
Hence, removing the size of the headers of our 453,200 extracted blocks (\mbox{$\sim$ 42} MB) 
from the total size of the blockchain at 2017/02/15, we obtain \mbox{$\sim$ 102} GB of transactions.
From this we conclude that \opreturn transactions consume
$\sim$0.3\% of the total space on the blockchain.

\begin{figure}
	\hspace{-5pt}
	\begin{tikzpicture}
	[
	pie chart,
	slice type={Assets}{blu},
	slice type={Notary}{verde},
	slice type={Digital Arts}{arancione},
	slice type={Other}{rosso},
	slice type={Empty}{viola},
	slice type={Unknown}{azzurro},
	pie values/.style={font={\small}},
	scale=2
	]
	
	\pie{}{26.7/Assets, 8.3/Notary, 32.8/Unknown, 10.8/Other, 15.7/Empty, 5.5/Digital Arts}
		
	\legend[shift={(0.5cm, 0.6cm)}]{{Assets}/Assets, {Notary}/Notary}
	\legend[shift={(0.5cm, 0cm)}]{{Digital Arts}/Digital Arts, {Other}/Other}
	\legend[shift={(0.5cm, -0.6cm)}]{{Empty}/Empty, {Unknown}/Unknown}
	
	\end{tikzpicture}
\caption{Distribution of transactions by category.}
\label{fig:GroupsPye}
\end{figure}
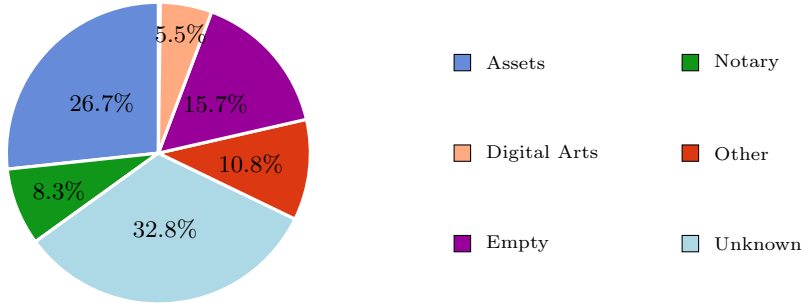

\subsection{Distribution of protocols by category}
\label{subsec:ResultCategories}

\Cref{fig:GroupsPye} displays how the \opreturn transactions are distributed
in the categories identified in \Cref{sec:qualitative}.
We note a relevant component of \textit{empty} and \textit{unknown} transactions. 
Although \textit{assets} protocols produce the highest number of transactions, 
the most numerous category is \textit{document notary}.

\Cref{fig:Groups} and the fourth column of \Cref{fig:table} suggest that,
originally, the protocols using \opreturn were in the categories \textit{assets} and \textit{notary}, 
while the other use cases were introduced subsequently 
(indeed, the \textit{other} category was not inhabited before the end of 2014).

Empty transactions use \opreturn without any data attached, so they are not associated to any protocol. 
We evaluate that $\sim 96\%$ of these transactions are related to the
transaction peaks discussed in~\Cref{subsec:ResultPeaks}. 
Since those peaks happened in the same period
of the stress tests and spam campaign discussed in~\cite{baqerstressing}, 
we conjecture that \textit{empty} transactions are related to those events%
\footnote{
To verify this conjecture we would need to compare the transaction identifiers of our 
\textit{empty} transactions with the identifiers of~\cite{baqerstressing}, 
which are not available online.
}.

The \textit{unknown} category contains $\sim 33\%$ of the \opreturn transactions.
We identify 3 protocols~\cite{ChainpointWebsite,DiplomaWebsite,counterparty} 
that write \opreturn data only as \textit{unknown} transactions.
We also identify one protocol~\cite{LaPreuveWebsite}
that besides using an identifier for saving document hashes, 
allows to save text messages without any identifier.

%% file: conclusions.tex
\section{Conclusions}

Our analysis shows an increasing interest in the \opreturn instruction. 
While in the first year of existence of \opreturn transactions (from March 2014) 
only a few hundreds of these transactions were appended per week, 
their usage has been steadily increasing since March 2015. 
In the last weeks of our experiments (February 2017) 
we counted $\sim$25,000 new \opreturn transactions per week, on average.
Overall, we estimate that 
\opreturn transactions constitute \mbox{$\sim 1\%$} of the transactions in the
blockchain, and use \mbox{$\sim 0.3\%$} of its space.

Besides using \opreturn and \emph{Pay-to-PubkeyHash} as shown in~\Cref{sec:background},
there are other techniques to save metadata on the Bitcoin blockchain.
With a slightly different flavour, 
the ``sign-to-contract'' and ``pay-to-contract''~\cite{Sign-to-contract,OpReturnAlternatives}
allow to prove that, if a certain transaction is redeemed,
then a certain value was known at the time it was put on the blockchain.
A benefit of these techniques is that they do not affect the size of transactions.
Comparing different methods to store metadata on Bitcoin
would be an interesting topic for future research.

Although the official Bitcoin documentation discourages the use
of the block\-chain to store arbitrary data%
\footnote{
The release notes of Bitcoin Core version 0.9.0 state that:
\emph{``Storing arbitrary data in the blockchain is still a bad idea; it is less costly and far more efficient to store non-currency data elsewhere.''}
},
the trend seems to be a growth in the number of blockchain-based applications 
that embed their metadata in \opreturn transactions.
We think that the main motivation for not using cheaper and more efficient storage
is the perceived sense of security and persistence of the Bitcoin blockchain.
If this trend will be confirmed,
the specific needs of these applications could affect
the future evolution of the Bitcoin protocol.

\paragraph{Related work.}

Besides ours, other projects aim at analysing metadata in the Bitcoin block\-chain.
For instance, \href{http://blockchainarchaeology.com}{blockchainarchaeology.com} 
collects files hidden in the blockchain.
These files are usually split into several parts, 
stored \eg on different output scripts in a transaction.
Various \href{https://github.com/spooktheducks}{techniques} are used to detect how the files 
were encoded (\eg by binary grep on the PNG pattern) and to reconstruct them.
The Bitcoin wiki~\cite{BitcoinOpReturnDescription} provides a list of protocols using \opreturn,
together with their identifiers.
Excluding those protocol identifiers that, at time of writing, are not used yet in any \opreturn transaction, 
the collection in~\cite{BitcoinOpReturnDescription} is strictly included in ours.
The website \href{http://opreturn.org/}{opreturn.org} shows charts about \opreturn transactions over time, 
organised by protocol,
and statistics about their usage on the last week and over the last two years.
The website \href{https://www.smartbit.com.au/op-returns}{smartbit.com} 
recognises some \opreturn identifiers and shows related statistics.
Finally, the website \href{https://www.kaiko.com/}{kaiko.com} sells data about Bitcoin, 
including data related to \opreturn transactions.

%% file: ack.tex
\paragraph{Acknowledgments.}

The authors thank the anonymous reviewers of  
\href{http://fc17.ifca.ai/bitcoin/}{BITCOIN 2017}
for their insightful comments on a preliminary version of this paper.
This work is partially supported by
Aut.\ Reg.\ of Sardinia P.I.A.\ 2013 ``NOMAD''.

%% file: appendixA.tex
\begin{figure}[h]
  \centering
  \begin{minipage}{0.5\textwidth}
    \centering
    \includegraphics[width=\textwidth]{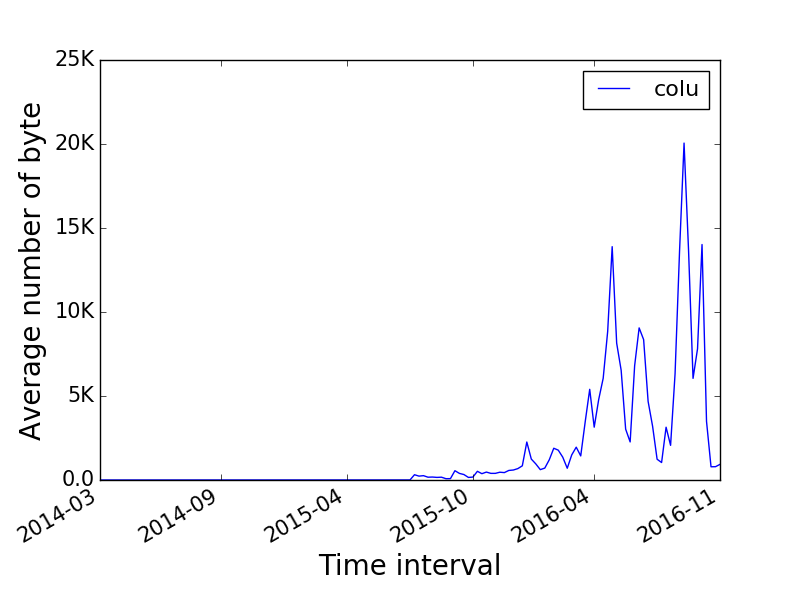}
    \label{fig:TransactionsColu}
  \end{minipage}\hfill
  \begin{minipage}{0.5\textwidth}
    \centering
    \includegraphics[width=\textwidth]{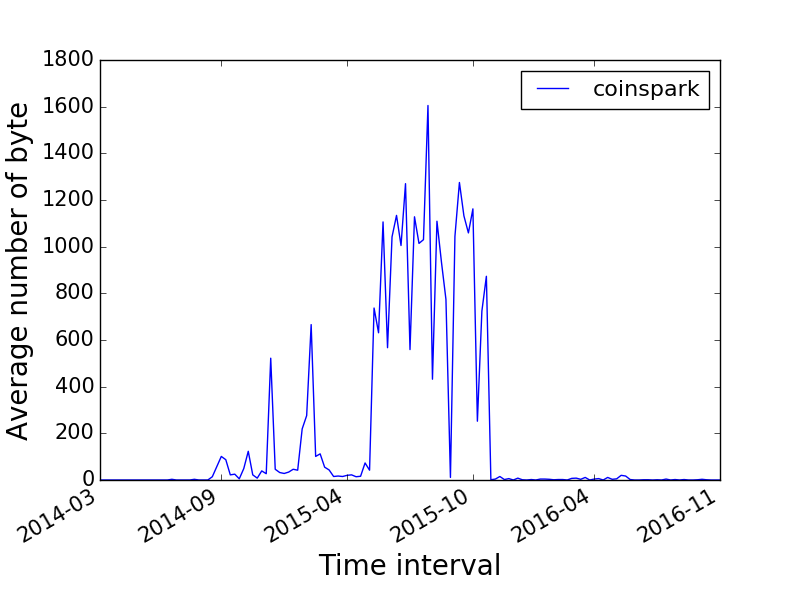}
    \label{fig:TransactionsCoinSpark}
  \end{minipage}

  \begin{minipage}{0.5\textwidth}
    \centering
    \includegraphics[width=\textwidth]{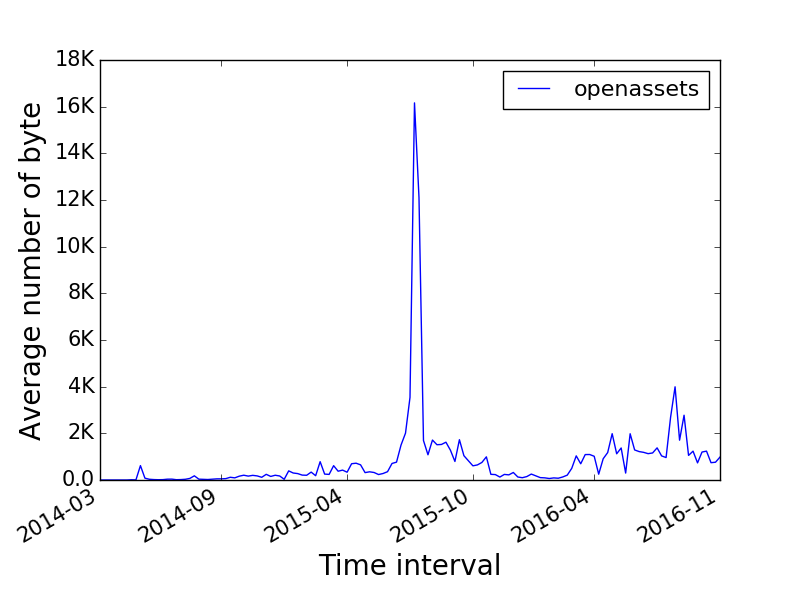}
    \label{fig:TransactionsOpenAssets}
  \end{minipage}\hfill
  \begin{minipage}{0.5\textwidth}
    \centering
    \includegraphics[width=\textwidth]{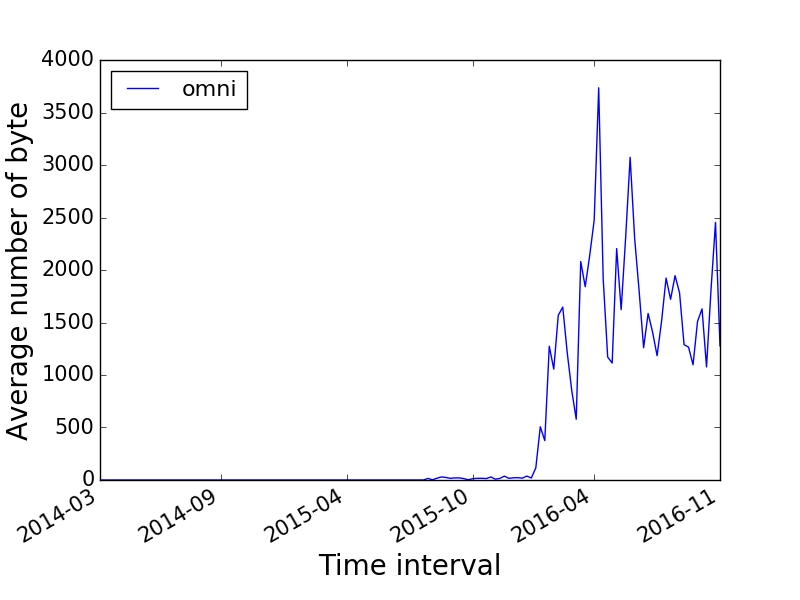}
    \label{fig:TransactionsOmni}
  \end{minipage}
  \caption{Assets charts.}
\end{figure}

%% file: appendixB.tex
\begin{figure}[t!]
  \centering
  \begin{minipage}{0.5\textwidth}
    \centering
    \includegraphics[width=\textwidth]{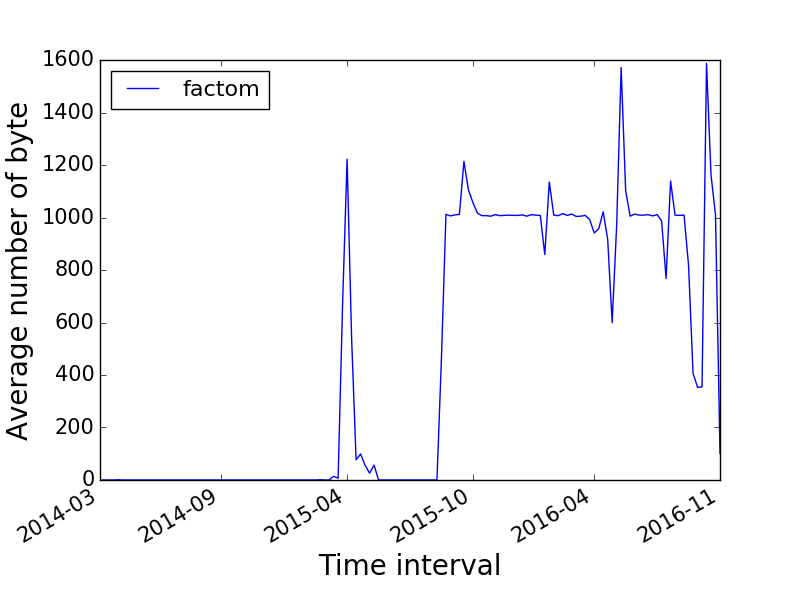}
    \label{fig:TransactionsFactom}
  \end{minipage}\hfill
  \begin{minipage}{0.5\textwidth}
    \centering
    \includegraphics[width=\textwidth]{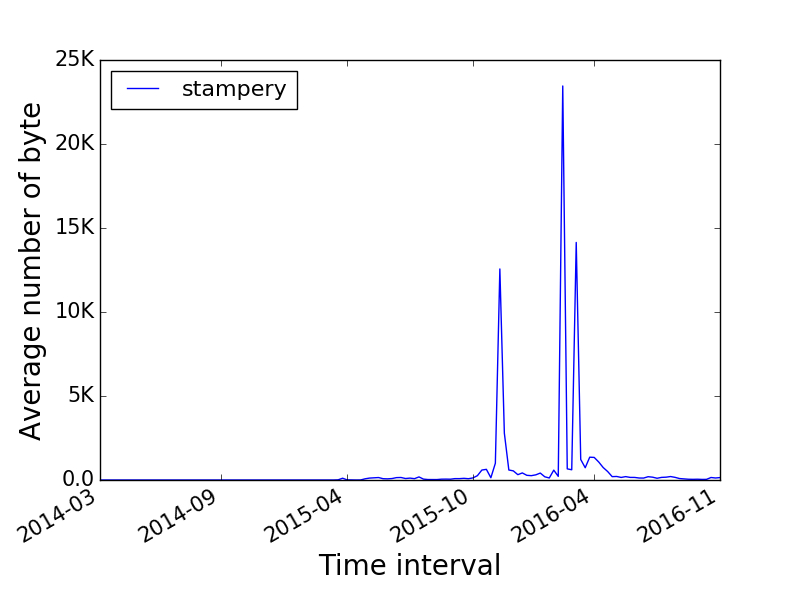}
    \label{fig:TransactionsStampery}
  \end{minipage}

  \centering
  \begin{minipage}{0.5\textwidth}
    \centering
    \includegraphics[width=\textwidth]{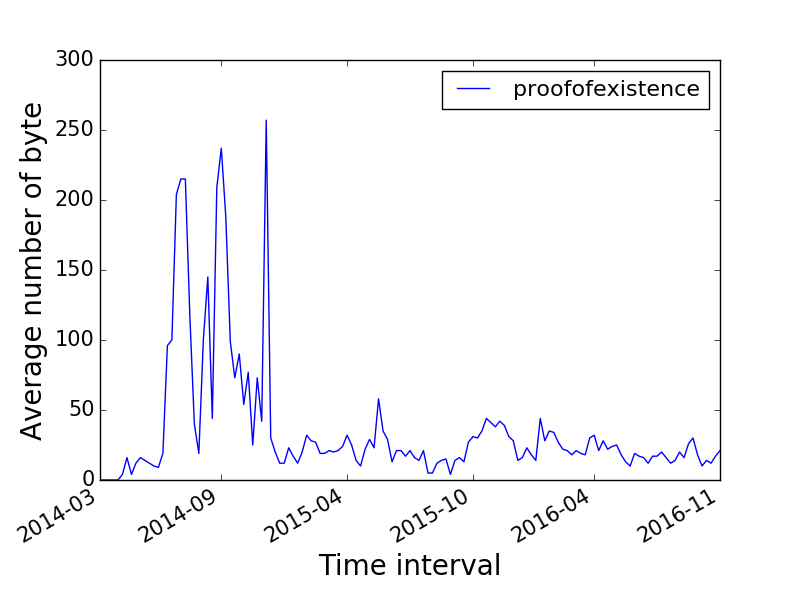}
    \label{fig:TransactionsProofOfExistence}
  \end{minipage}\hfill
  \begin{minipage}{0.5\textwidth}
    \centering
    \includegraphics[width=\textwidth]{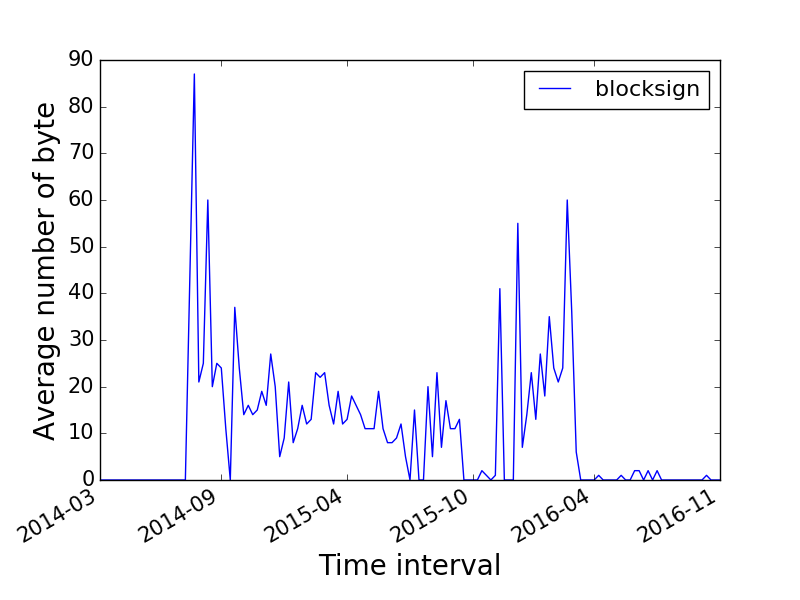}
    \label{fig:TransactionsBlocksign}
  \end{minipage}

  \centering
  \begin{minipage}{0.5\textwidth}
    \centering
    \includegraphics[width=\textwidth]{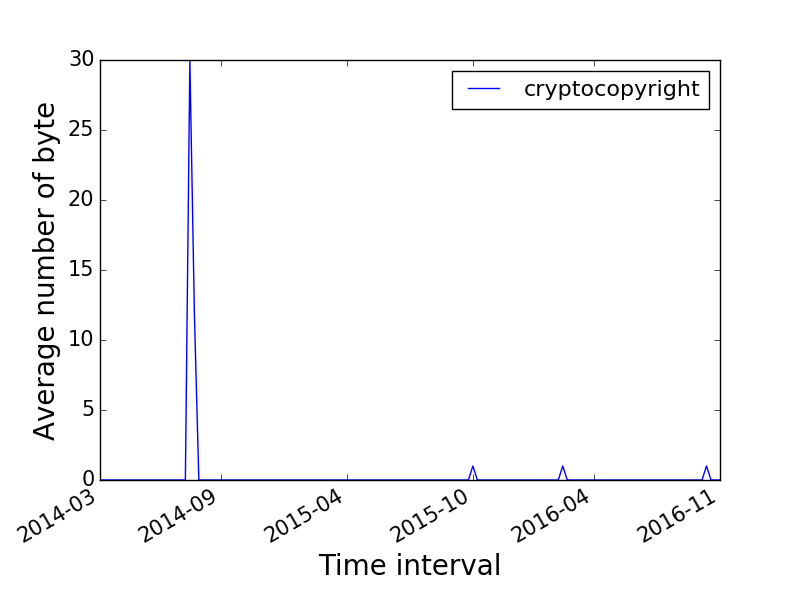}
    \label{fig:TransactionsCryptoCopyright}
  \end{minipage}\hfill
  \begin{minipage}{0.5\textwidth}
    \centering
    \includegraphics[width=\textwidth]{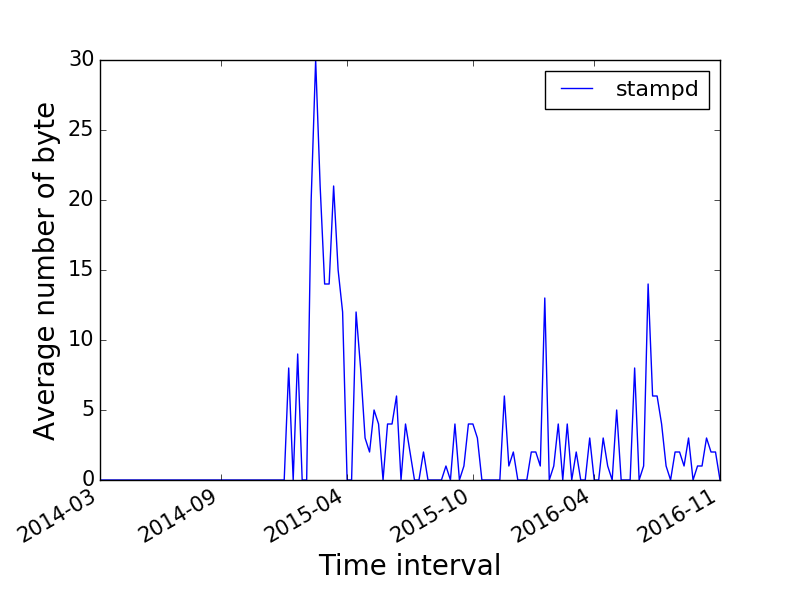}
    \label{fig:TransactionsStampd}
  \end{minipage}
  \caption{Document Notary charts (1).}
\end{figure}

\begin{figure}[t!]
  \centering
  \begin{minipage}{0.5\textwidth}
    \centering
    \includegraphics[width=\textwidth]{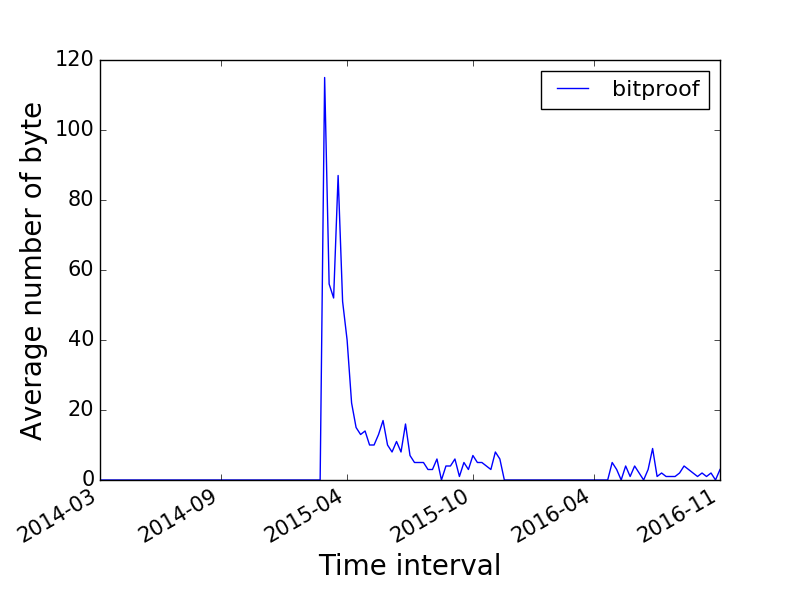}
    \label{fig:TransactionsBitproof}
  \end{minipage}\hfill
  \begin{minipage}{0.5\textwidth}
    \centering
    \includegraphics[width=\textwidth]{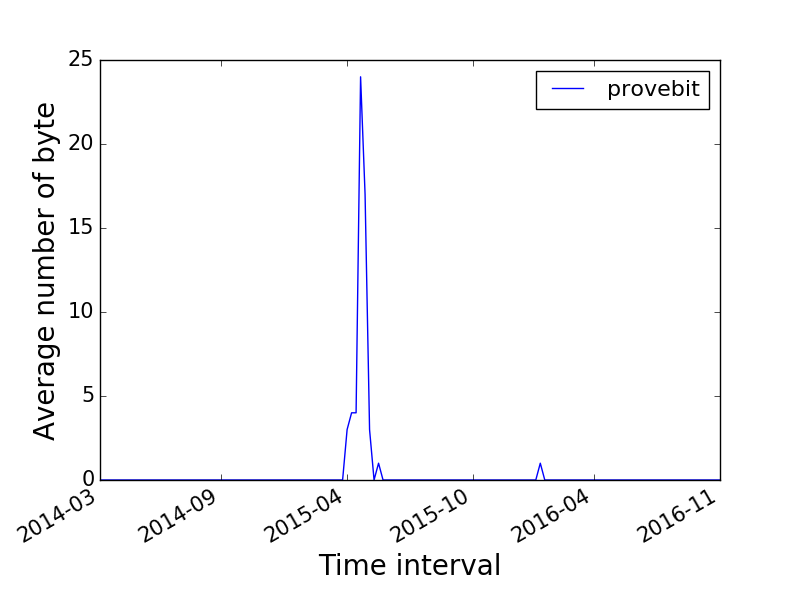}
    \label{fig:TransactionsProveBit}
  \end{minipage}

  \centering
  \begin{minipage}{0.5\textwidth}
    \centering
    \includegraphics[width=\textwidth]{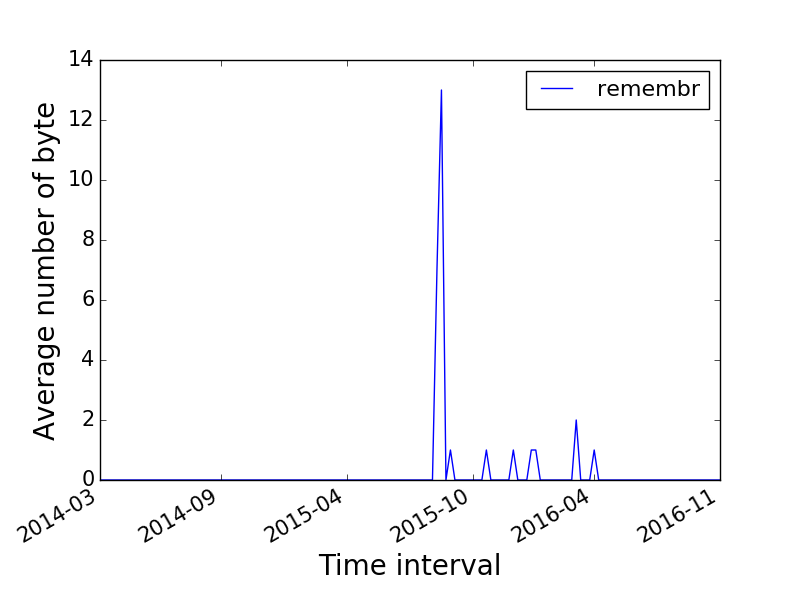}
    \label{fig:TransactionsRemembr}
  \end{minipage}\hfill
  \begin{minipage}{0.5\textwidth}
    \centering
    \includegraphics[width=\textwidth]{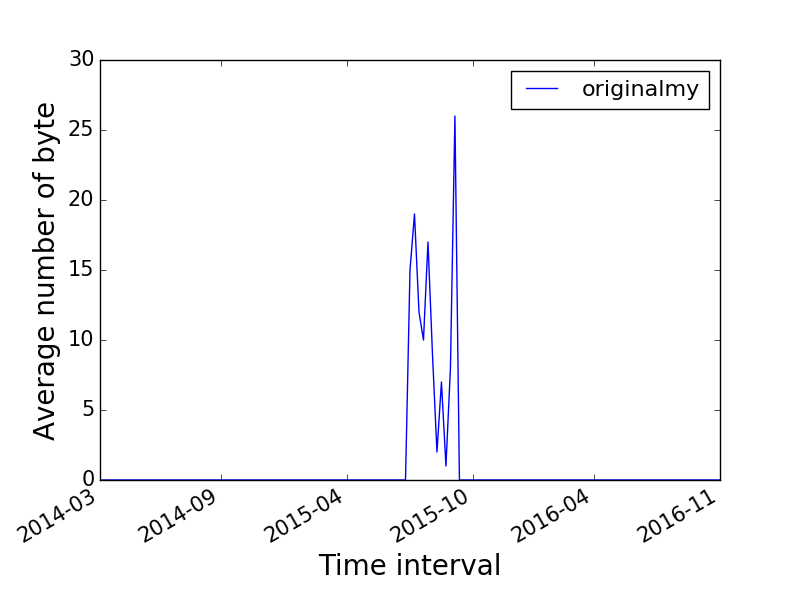}
    \label{fig:TransactionsOriginalMy}
  \end{minipage}

  \centering
  \begin{minipage}{0.5\textwidth}
    \centering
    \includegraphics[width=\textwidth]{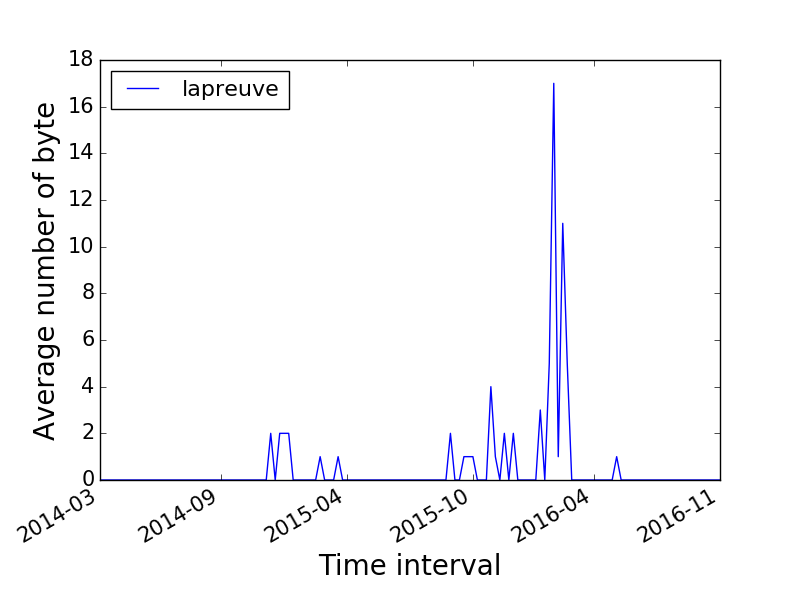}
    \label{fig:TransactionsLaPreuve}
  \end{minipage}\hfill
  \begin{minipage}{0.5\textwidth}
    \centering
    \includegraphics[width=\textwidth]{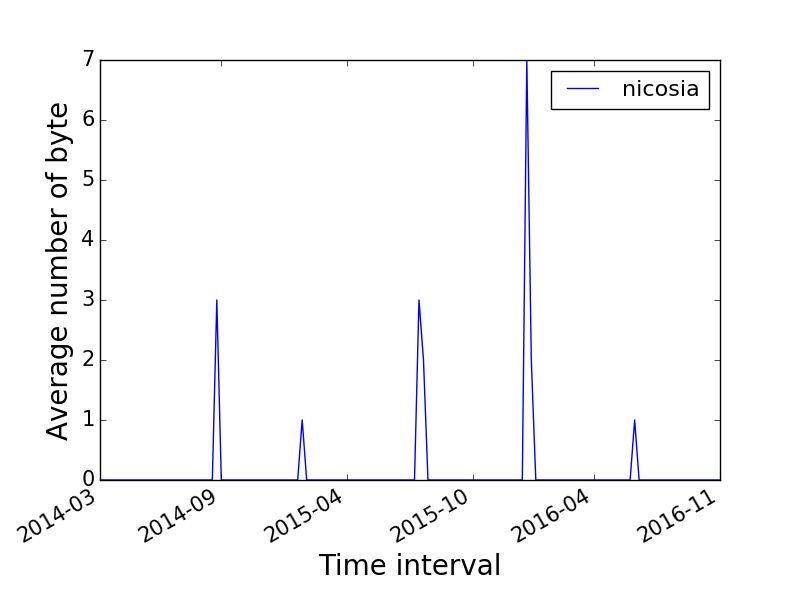}
    \label{fig:TransactionsNicosia}
  \end{minipage}
  \caption{Document Notary charts (2).}
\end{figure}

%% file: appendixC.tex
\begin{figure}[h]
	\centering
	\begin{minipage}{0.5\textwidth}
		\centering
		\includegraphics[width=\textwidth]{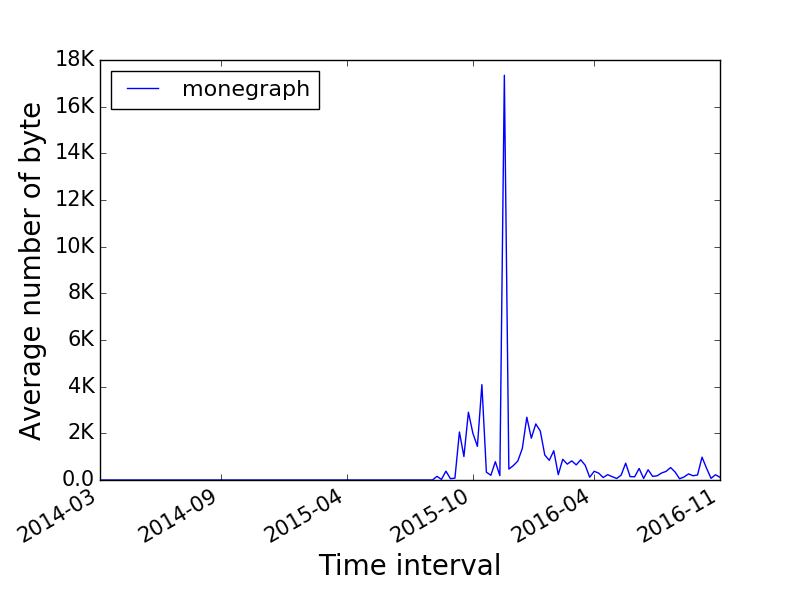}
		\label{fig:Monegraph}
	\end{minipage}\hfill
	\begin{minipage}{0.5\textwidth}
		\centering
		\includegraphics[width=\textwidth]{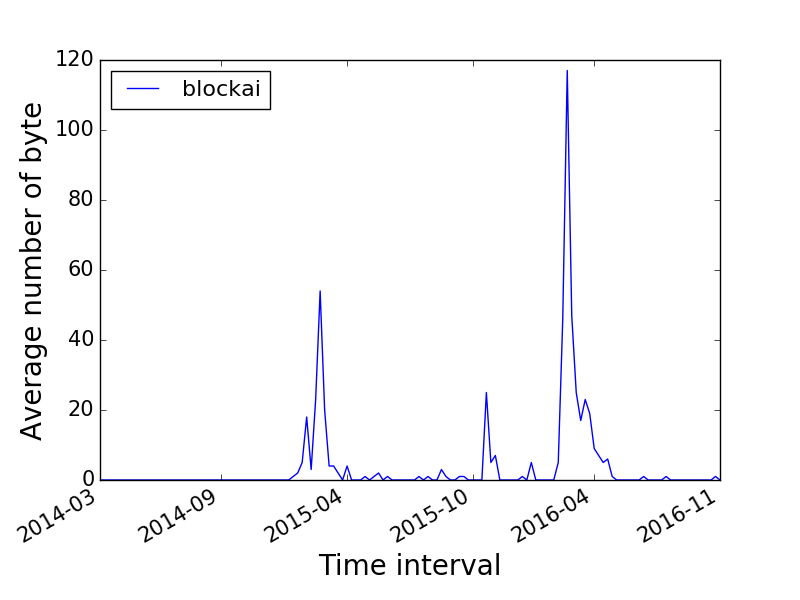}
		\label{fig:Blockai}
	\end{minipage}

	\centering
	\includegraphics[width=0.5\textwidth]{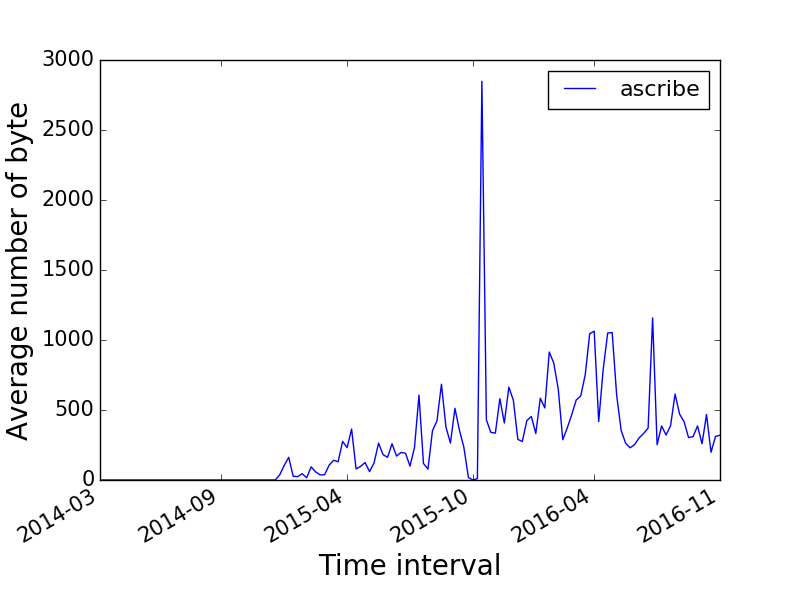}
        \caption{Digital arts charts.}
	\label{fig:Ascribe}
\end{figure}

%% file: appendixD.tex
\begin{figure}[h]
  \centering
  \begin{minipage}{0.5\textwidth}
    \centering
    \includegraphics[width=\textwidth]{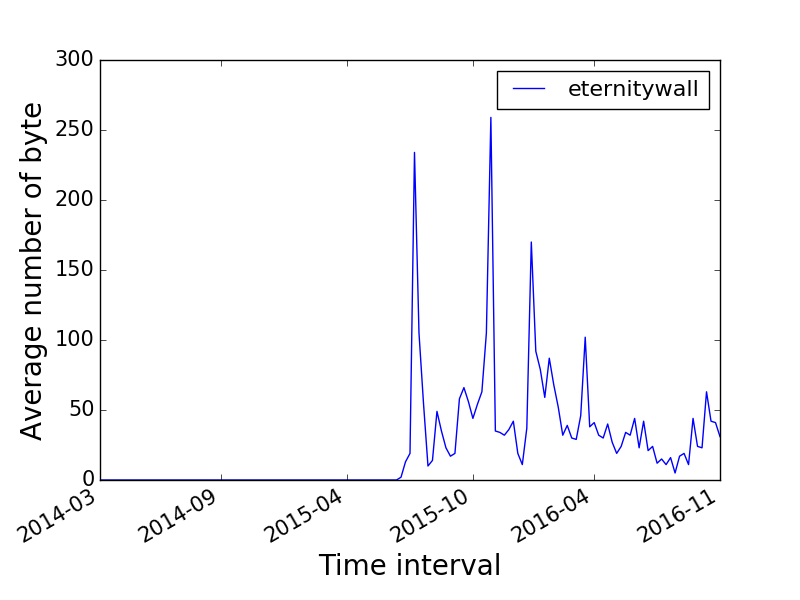}
    \label{fig:EternityWall}
  \end{minipage}\hfill
  \begin{minipage}{0.5\textwidth}
    \centering
    \includegraphics[width=\textwidth]{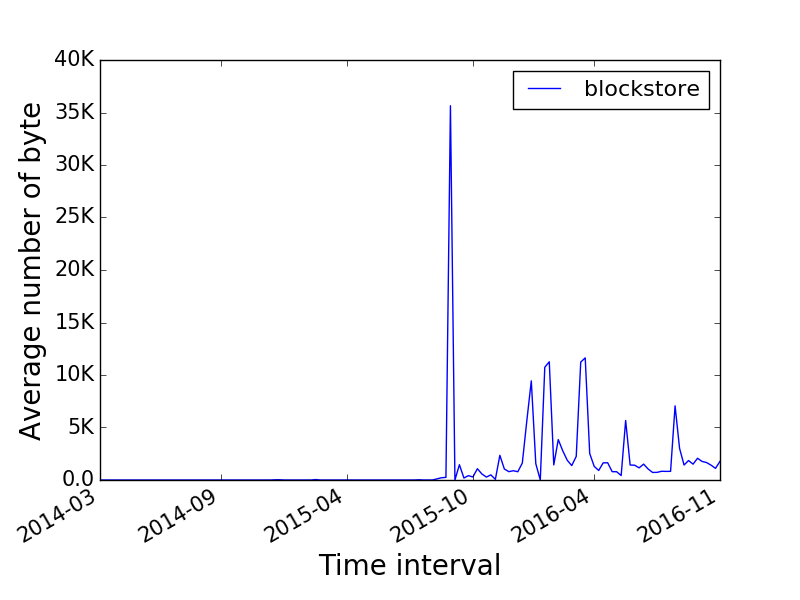}
    \label{fig:Blockstore}
  \end{minipage}

  \centering
  \begin{minipage}{0.5\textwidth}
    \centering
    \includegraphics[width=\textwidth]{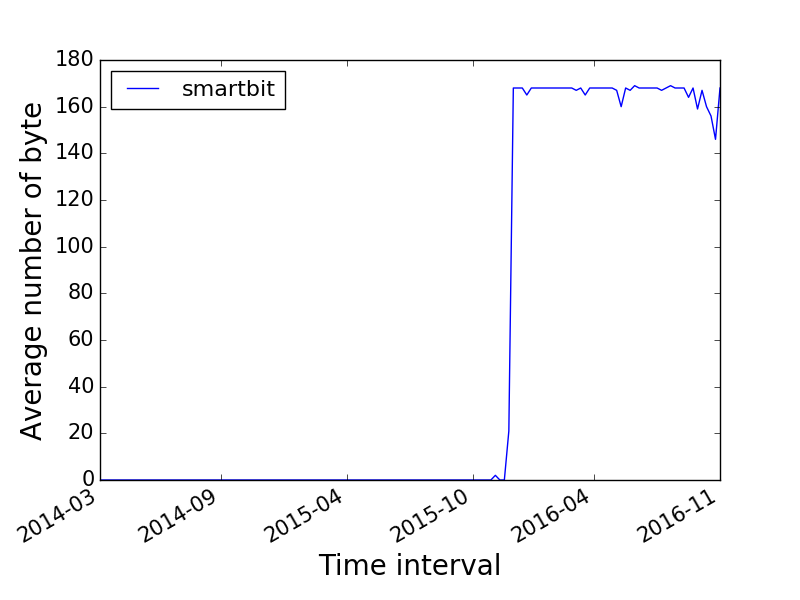}
    \label{fig:Smartbit}
  \end{minipage}\hfill
  \caption{Other Charts}
\end{figure}